# Magneto-Rayleigh-Taylor instability driven by a rotating magnetic field: Cylindrical liner configuration

Shu-Chao Duan*, Long Yang, Bo Xiao, Ming-Xian Kan, Gang-Hua Wang, and Wei-Ping Xie*
Institute of Fluid Physics, China Academy of Engineering Physics (CAEP),
Mianyang 621999, China

*Abstract*—We propose using a directional time-varying (rotating) driving magnetic field to suppress magneto-Rayleigh-Taylor (MRT) instability in dynamic Z-pinches. A rotational drive magnetic field is equivalent to two magnetic-field components, Θ and Z, that alternate in time, referred to as an alternate Theta-Z-pinch configuration. We consider the finitely thick cylindrical liner configuration in this paper. We numerically integrate the perturbation equation to stagnation time based on the optimal background unperturbed trajectories. We assess the cumulative growth of the dominant mode selected by some mechanism at the beginning of an implosion. The maximum e-folding number at stagnation of the dominant mode of an optimized alternate Theta-Z-pinch is significantly lower than that of the standard Theta- or Z-pinch. The directional rotation of the magnetic field contributes to suppress the instabilities, independent of the finite thickness. The finite thickness effect plays a role only when the orientation of the magnetic field varies in time whereas it does not appear in the standard Theta- or Z-pinch. The rotating frequency of the magnetic field and the thickness of liner are both having a monotonic effect on suppression. Their synergistic effect can enhance the suppression on MRT instability. Because the MRT instability can be well suppressed in this way, the alternate Theta-Z-pinch configuration has potential applications in liner inertial fusion.\*

## I. Introduction

Using the equilibrium Z-pinch as a magnetic-confinement fusion system was previously found to be difficult because of the unavoidable exchange or quasi-exchange mode instability. In modern times, with the development of pulsed-power technology and the emergence of the dynamic Z-pinch, the fusion field has renewed its interest in the Z-pinch [1-6]. In a dynamic Z-pinch system, magneto-Rayleigh–Taylor (MRT) instability [7-10] is inevitable from implosions of the plasma driven by magnetic pressure. While developing along with other magnetohydrodynamic modes of instability, the MRT instability grows much faster and by undermining the pinch symmetry is the most dangerous.

We considered the use of a directional time-varying (rotating) driving magnetic field to suppress MRT instability in dynamic Z-pinches [11-44]. We extend the work of [14] to finitely thick cylindrical liners in this paper. We numerically integrate the perturbation equation (Sec. II) to stagnation time based on the optimal background unperturbed trajectories (Sec. III). We assess the cumulative growth of the dominant mode selected by some mechanism at the beginning of an implosion (in Sec. IV).

## II. Model for the Perturbations

We consider a finitely thick cylindrical liner configuration (Fig. 1). Magnetic fields with time-varying orientations are present in the inner and outer surfaces of the liner. The liner is assumed perfectly conducting, so the magnetic field is discontinuous and must remain parallel to the surface. In the reference frame co-moving with the liner, there is an effective gravitational field directed in the $r$ direction. The liner is assumed to be incompressible, under which the density and thickness cannot be simultaneously kept constant during implosion of the liner: one parameter may remain constant with time while the other varies. In real compressible conditions, both density and thickness evolve with time during an implosion; however, this study ignores this constraint, assuming that both density and thickness are constant.

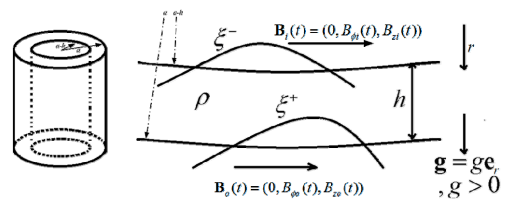

Fig. 1. Magneto-Rayleigh-Taylor model of a perfectly conducting incompressible liner accelerated in the $-r$ direction by pressure differences between the outer and inner magnetic fields.

The linearized movement equations are

This work is supported by the NSFC (Grant Nos. 11405167, 51407171, 11571293, 11605188, and 11605189) and the Foundation of the China Academy of Engineering Physics (No. 2015B0201023).

$$\rho \ddot{\xi}^{\pm} - \rho g \partial_r \xi^{\pm} + \partial_r \delta p_m^{\pm} = 0 \ . \quad (1)$$

The magnetic-pressure perturbation can be found by using the condition that the magnetic field remains parallel to the perturbed surface. We assume that the perturbed velocity field is irrotational. After some derivation we can get

$$(\ddot{\xi} - D_+ g k \xi)(\ddot{\xi} - D_- g k \xi) = 0 \ . \quad (2)$$

It can be proved that $D_-$ is always negative and thus related to the oscillatory modes. These modes are assumed to be ultimately damped by some dissipation mechanism, so the $D_-$ branch is deserted not to considered further; only the $D_+$ branch is considered below, along with solutions to

$$\ddot{\xi} - D_+ g k \xi = 0 \ . \quad (3)$$

This model recovers several of the results of [15-17] in the shell or slab limit.

### III. Background Unperturbed Trajectories

Perturbations' evolution is riding on the background implosion trajectories of dynamic pinches. A rotational drive magnetic field is equivalent to two magnetic-field components, Θ and Z, that alternate in time, referred to as an alternate Theta-Z-pinch configuration. The normalized kinetic equation of implosion of the alternate Theta-Z-pinch in the zero-dimensional shell model is

$$\ddot{a} = -\Pi(I_z^2/a + \alpha a I_\phi^2) \ . \quad (4)$$

We fixed the total drive current to a sine waveform. Two ways of decomposing the total current into the Θ- and Z-direction drive currents are discussed:

$$I_t = \sin(t),\ I_z = I_t \cos(t + td),\ I_\theta = I_t \sin(t + td)\ , \quad (5)$$

$$I_z = I_\theta = \sin(t) \ . \quad (6)$$

The standard Z- and Theta-pinches can be recovered as the limits of these types of driving.

We fixed implosions to the optimal trajectories because an actual dynamic pinch is generally expected to run near the optimal trajectory.

### IV. Results

We numerically integrate the perturbation equation to stagnation time based on the optimal background unperturbed trajectories. Finally, we assess the maximum cumulative growth, e-folding number, of the dominant mode with millimeter-scale [18-26] selected by some mechanism at the beginning of an implosion [27].

The results in the case of driving described by Eq. (5) are shown in Fig. 2. The maximum e-folding number at stagnation of the dominant mode of an optimized alternate Theta-Z-pinch is significantly lower than that of the standard Theta- or Z-pinch.

For classical cases where the direction of the magnetic field is fixed with time, the finite thickness does not suppress the perpendicular mode, although suppresses the non-perpendicular mode. Overall, for the fastest (perpendicular) mode, finite thickness has no effect. In Fig. 2, the weak variation on thickness in the standard Theta- or Z-pinch is a geometric—not finite thickness—effect.

For the time-varying (rotating) magnetic-field orientation cases, each mode can be on average subjected to a finite thickness suppression with time. Even when the thickness is zero, the directional time-varying of the magnetic field still suppresses the MRT instability. Therefore, the directional time-variation of the magnetic field is independent of the finite thickness in terms of suppressing MRT instability, both thus cooperating to enhance the suppression.

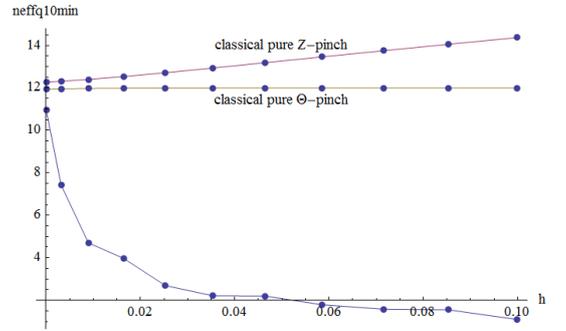

Fig. 2. Maximum e-folding numbers of the dominant mode at stagnation of the optimal trajectory driven by current described by Eq. (5) with a fixed $t_d$ =0.

For example, the time history of the magnetic-field orientation angle for the optimization point corresponding to $h$ =0.1 (Fig. 3) shows that the driving magnetic field is rotating during the implosion.

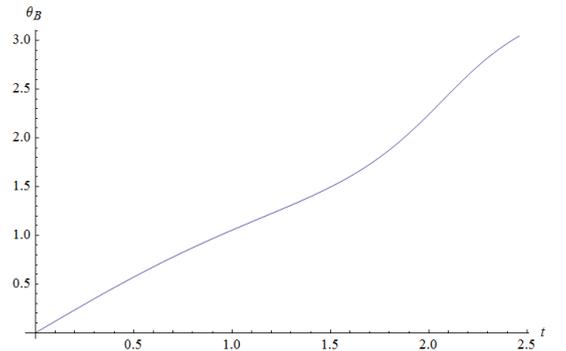

Fig. 3. Evolution of orientation angle of driving magnetic field for the point $h$=0.1 of Fig. 2.

The results in the case of driving described by Eq. (6) are shown in Fig. 4 and Fig. 5. Herein rotates slower the driving magnetic field than that in the preceding case driven by Eq. (5); accordingly, the degree of the suppression of MRT instability is smaller. This indicated that the suppression becomes stronger as the driving magnetic field rotates faster.

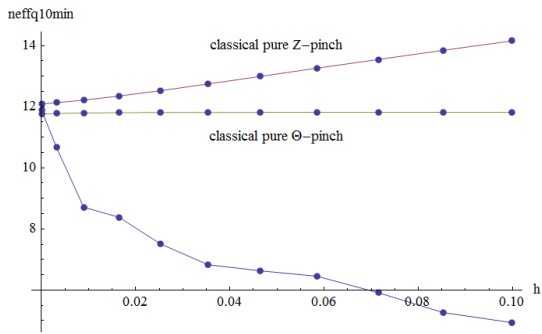

Fig. 4. Maximum e-folding numbers of the dominant mode at stagnation of the optimal trajectory driven by current described by Eq. (6).

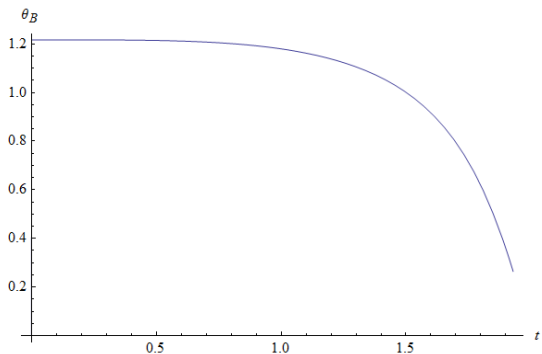

Fig. 5. Evolution of orientation angle of driving magnetic field for the point $h=0.1$ of Fig. 4.

## V. CONCLUSIONS

MRT instability can be well suppressed by the directional time-varying (rotating) driving magnetic field. The directional rotation of the magnetic field contributes to suppress the instabilities, independent of the finite thickness. The finite thickness effect plays a role only when the orientation of the magnetic field varies in time whereas it does not appear in the standard Theta- or Z-pinch. The rotating frequency of the magnetic field and the thickness of liner are both having a monotonic effect on suppression. Their synergistic effect can enhance the suppression on MRT instability. Because the MRT instability can be well suppressed in this way, the alternate Theta-Z-pinch configuration has potential applications in liner inertial fusion.

The dominant modes are not automatically generated by evolution in the present work, although they are determined via certain rationales. Because of the simplicity of our formula, viscous or surface tension mechanisms can be easily added to address this issue. On this basis, to assess the maximum growth of MRT instability would be more accurate. This work is in progress and is to be reported soon. Alternatively, one may employ magnetohydrodynamic simulations to solve this problem and include other effects. A comprehensive analysis of fusion applications using three-dimensional simulations remains an open problem.